\documentclass{article}
\usepackage{amsmath,graphicx,amssymb}
\usepackage[preprint]{spconf}
\copyrightnotice{\copyright\ IEEE 2023}
\toappear{To appear in {\it Proc.\ ICASSP2023,
                   June 04-10, 2023, Rhodes Island, Greece}}
\usepackage{booktabs}
\usepackage[hidelinks]{hyperref}
\hypersetup{
  colorlinks   = true, 
  urlcolor     = black, 
  linkcolor    = blue, 
  citecolor   = green 
}
\usepackage[bibstyle=ieee, citestyle=numeric-comp, firstinits, url=false, doi=false, maxbibnames=50]{biblatex}
\bibliography{refs}

\title{Spectral Clustering-aware Learning of Embeddings\\ for Speaker Diarisation}
\name{Evonne P.C. Lee, Guangzhi Sun, Chao Zhang, Philip C. Woodland\thanks{Guangzhi Sun is supported by Cambridge Trust}}
\address{Cambridge University Engineering Dept., Trumpington St., Cambridge, CB2 1PZ U.K.\\
\small{\texttt{\{epcl2,gs534,cz277,pcw\}@eng.cam.ac.uk}}}
\begin{document}
\ninept
\maketitle
\begin{abstract}
In speaker diarisation, speaker embedding extraction models often suffer from the mismatch between their training loss functions and the speaker clustering method. 
In this paper, we propose the method of spectral clustering-aware learning of embeddings (SCALE) to address the mismatch. 
Specifically, besides an angular prototypical (AP) loss, SCALE uses a novel affinity matrix loss which directly minimises the error between the affinity matrix estimated from speaker embeddings and the reference.
SCALE also includes $p$-percentile thresholding and Gaussian blur as two important hyper-parameters for spectral clustering in training.
Experiments on the AMI dataset showed that speaker embeddings obtained with SCALE achieved over 50\% relative speaker error rate reductions using oracle segmentation, and over 30\% relative diarisation error rate reductions using automatic segmentation when compared to a strong baseline with the AP-loss-based speaker embeddings. 


\end{abstract}
\section{Introduction}
\label{sec:intro}
Speaker diarisation is the task of finding `who spoke when' in an audio stream with multiple speakers. 
It has many applications including speech recognition and information retrieval \textit{etc}. 
In a typical diarisation pipeline, audio data that contains speech is divided into fixed-length segments (e.g. 2 seconds), also known as windows. A speaker embedding is then extracted for each window using deep neural networks trained for speaker classification, known as d-vectors \cite{d_vector, sv_dvec, sd_lstm, sd_dnn, dnn_tisv}, and a speaker label is assigned via clustering algorithms. In particular, spectral clustering \cite{spectral_clustering} has been widely adopted in many recent diarisation systems \cite{sc_diar, gs, sc_auto}, which identifies groups of nodes based on the graph affinity matrix computed from speaker embeddings in a fully unsupervised process.

Using neural network-derived d-vectors in a diarisation system often suffers from the mismatch between training and clustering since d-vectors are not trained to discriminate the relative speaker differences across multiple utterances which is particularly important for unsupervised clustering. 
To mitigate such a mismatch, metric learning losses such as the angular prototypical (AP) loss \cite{metric_srec, sv_ap, sv_ap_amp} and the angular margin prototypical (AMP) loss \cite{sv_ap_amp} have been applied when fine-tuning a d-vector extraction model for speaker verification. 
Besides, work has been performed on supervised clustering methods \cite{e2e_sd_sa, dnc}, or adaptive hyper-parameters for spectral clustering have been adopted \cite{turntodiar}.
Recently, representations from pre-trained speech foundation models \cite{w2v, w2v2, hubert}, such as wav2vec 2.0, have achieved superior performance on speaker diarisation \cite{superb,tony_is22}. However, a similar mismatch still exists since the pre-trained models are often fine-tuned by speaker classification \cite{tony_is22, zhang2021beijing}. 

To resolve the mismatch between the speaker embedding extraction model and the downstream unsupervised speaker clustering, in this paper, we propose the spectral clustering-aware learning of embeddings (SCALE)\footnote{Code available at \url{https://github.com/epcl2/scale}} method for fine-tuning the pre-trained representations for speaker diarisation.
Specifically, an AP loss is used to fine-tune the wav2vec 2.0 model to encourage the representations to discriminate the relative speaker differences across multiple utterances. 
To simulate spectral clustering during training, an extra affinity matrix (AM) loss is used, which minimises the mean squared errors between the reference and hypothesis affinity matrices and helps learn the structure of the speaker embedding space.
Furthermore, the two key steps in spectral clustering, namely the Gaussian blurring and $p$-th percentile thresholding, are also accounted for during training in both the AP loss and the AM loss.
Speaker diarisation experiments were performed on the widely used AMI meeting corpus \cite{ami}, as well as a combination of AMI and VoxCeleb1$+$2 \cite{vox1, vox2} datasets. Consistent and statistically significant improvements were achieved using SCALE on both settings
compared to both the speaker classification baseline and a strong AP loss baseline.

The rest of the paper is organised as follows. Section~\ref{sec:rwork} gives an overview of related work. Section~\ref{sec:cl} explains about the speaker diarisation system. In Section~\ref{sec:scale}, SCALE is presented. The experimental setup is given in Section \ref{sec:exp} and the results are presented in Section~\ref{sec:res}.

\section{Related Work}
\label{sec:rwork}

\subsection{Wav2vec 2.0 for speech tasks}
Wav2vec 2.0 \cite{w2v2} learns generic speech representations via self-supervised learning on large amounts of unlabelled speech data.
 It is trained to identify the correct quantised latent representation from a set of distractors, where the distractors are sampled uniformly from other masked time steps of the same utterance.
Despite its success in various speech tasks, fine-tuning the model has only achieved moderate performance on speaker-related tasks, 
including speaker recognition \cite{w2v2_srec} and verification \cite{fan2020exploring}, and speaker diarisation \cite{superb,tony_is22}, due to the mismatch between training and inference \cite{ssl_asv, unispeech_sat}.

\subsection{Loss functions for speaker diarisation}
For speaker-related tasks,  metric-learning-based loss functions have demonstrated competitive results \cite{metric_srec}. Due to the constraints from the unsupervised clustering present later in the pipeline, it is preferable to extract speaker embeddings with small intra-speaker distances and large inter-speaker distances. Work in \cite{triplet} proposed the triplet loss which required a triplet to be selected meticulously as the performance relied on the ``difficulty'' of the negative exemplars. The prototypical loss was proposed in \cite{prototypical}, where a query embedding was pushed away from the centroid of all negative samples based on the squared Euclidean distance in a mini-batch.
Instead of using only one utterance from each speaker as the query, in the generalised end-to-end loss \cite{gene2e}, every utterance in the mini-batch served as a query. 
Similar to the prototypical loss, the angular prototypical (AP) loss \cite{metric_srec} used only one utterance from each class as the query, with a cosine similarity-based metric. Work in \cite{cssl} used a contrastive self-supervised learning approach for text-independent speaker verification, where they adopted the AP loss. 

\section{Speaker Diarisation Pipeline}
\label{sec:cl}

The full speaker diarisation pipeline in this paper comprises several stages, including voice activity detection (VAD), change point detection (CPD), speaker embedding extraction and the spectral clustering stage. Neural VAD and CPD were used, which followed \cite{gs}. The VAD detects audio that contains speech and CPD splits speech segments into speaker-homogeneous segments. Each speaker-homogeneous segment is split into multiple windows of the same length, and speaker embeddings are extracted for each window using a fine-tuned wav2vec 2.0 encoder. Finally, spectral clustering is performed at the window level and all windows in the same speaker-homogeneous segment are assigned to the same speaker.

\subsection{Fine-tuning with angular prototypical loss}
\label{ssec:ft}
The wav2vec 2.0 encoder is fine-tuned with an extra output layer based on the AP loss.
For each mini-batch, $N$ utterances, $\mathbf{u}^\text{a}_{1},\ldots, \mathbf{u}^\text{a}_{N}$ from $N$ distinct speakers are randomly selected and act as the anchor utterances. Another $N$ utterances, $\mathbf{u}^\text{p}_{1},\ldots, \mathbf{u}^\text{p}_{N}$, are selected from the same $N$ speakers associated with the anchor utterances, which act as the positive utterances. 
For each pair of utterance $i$ ($1\leq i\leq N$),  $\mathbf{u}^\text{a}_{i}$ and  $\mathbf{u}^\text{p}_{i}$ have the same speaker identity but $\mathbf{u}^\text{a}_{i}\neq\mathbf{u}^\text{p}_{i}$. All anchor utterances from other speakers $j$ ($j \neq i$) in the same mini-batch serve as the negative samples for speaker $i$. The anchor and positive embeddings $\mathbf{e}^\text{a}_{i}$ and $\mathbf{e}^\text{p}_{i}$ are derived from the penultimate layer of the model (\textit{i.e.} the final wav2vec 2.0 encoder layer).
The AP loss, $\mathcal{L}_{\text{AP}}$, used to optimise the model is then 
 %
\begin{equation}
    \mathcal{L}_\text{AP} = - \frac{1}{N}\sum_{i=1}^N \log 
    \frac{\exp({\mathbf{S}_{i, i}})}
    {\sum_{j=1}^N \exp({\mathbf{S}_{i, j}})}
    \label{eq:ap}
\end{equation}
where $\mathbf{S}_{i, j}$ is an element in the similarity matrix $\mathbf{S}$, defined as the scaled cosine similarity between each pair of embeddings. That is,
\begin{equation}
    \mathbf{S}_{i, j} = 
    \begin{cases}
        w \cdot \textrm{sim}(\mathbf{e}^\text{a}_{i}, \mathbf{e}^\text{p}_{j})+b & \textrm{if }i = j \\
        w \cdot \textrm{sim}(\mathbf{e}^\text{a}_{i}, \mathbf{e}^\text{a}_{j})+b & \textrm{otherwise}
    \end{cases}
\end{equation}
where sim($\mathbf{x}, \mathbf{y}$) = $(\cos(\mathbf{x},\mathbf{y})+1)/2$, and $w$ and $b$ are trainable scalar values. 
In $\mathbf{S}=(\mathbf{S}_{i,j})\in\mathbb{R}^{N\times N}$, the similarity score of diagonal terms is computed between the anchor embedding and the positive embedding; the similarity score of the non-diagonal terms is computed between anchor embeddings to simulate the situation during spectral clustering. With the AP loss, the model explicitly minimises the distance among embeddings from the same speaker across different utterances (intra-speaker distances) and maximises the distance among embeddings from different speakers (inter-speaker distances). This differs from the pre-training of wav2vec 2.0 where the distractors were sampled from the same utterance. 

\subsection{Spectral clustering}
\label{sec:sc}
In spectral clustering, the affinity matrix $\mathbf{A}$ is first constructed, where $\mathbf{A}_{ij}$ is the cosine similarity between the embeddings of the $i$-th and $j$-th window for $i \neq j$. 
A modified version of spectral clustering based on \cite{sd_lstm} was used, where a sequence of refinement steps was applied to de-noise the affinity matrix. Two crucial steps, Gaussian blur and row-wise thresholding are highlighted here.

    \noindent
    \textbf{(i) Gaussian blur}: A Gaussian kernel with standard deviation $\sigma$ is used to smooth the data. Since window-level clustering is used, neighbouring embeddings are often derived from the same utterance, and hence should have similar values in the affinity matrix. Gaussian blur preserves this property among neighbouring windows. 

    \noindent
    \textbf{(ii) Row-wise thresholding}: Any element whose value ranked less than a particular row's $p$-th percentile is set to be zero, which ``zeroes out'' the affinities between embeddings from two distinct speakers. Ideally, the similarity scores of the embeddings belonging to different speakers should be below this threshold so that they can be ``thresholded-out'', whereas those of the  embeddings belonging to the same speakers should be above this threshold and reserved.

\section{Speaker Clustering with SCALE}
\label{sec:scale}

In Section \ref{sec:cl}, the model was fine-tuned without accounting for the clustering stage. 
To further reduce the mismatch between training and clustering, three steps are introduced to training, i.e. the AM loss, absolute thresholding and relative thresholding with Gaussian blurring. SCALE modifies the loss function such that the model is aware of the clustering stage and the associated refinement steps.

\subsection{Affinity matrix loss}
\label{mse_only}
The AM loss encourages the hypothesis affinity matrix $\mathbf{A}$, constructed using real speaker embeddings to be close to the reference affinity matrix with 1 for pairs from the same speaker and 0 for pairs from different speakers. 
By sampling $N$ pairs of utterances for $N$ distinct speakers in each mini-batch (as in Section~\ref{ssec:ft}), the reference affinity matrix is an identity matrix $\mathbf{I}$. $\mathcal{L}_\text{AM}$, the AM loss calculating the mean squared errors between $\mathbf{I}$ and $\mathbf{A}$, is
\begin{equation}
    \mathcal{L}_\text{AM} = \frac{1}{N^2}\sum_{i=1}^{N} \sum_{j=1}^{N}
    (\mathbf{I}_{i,j} - \mathbf{A}_{i,j})^2
\end{equation}
For $\mathcal{L}_\text{AM}$, $\mathbf{A}$ is equivalent to $\mathbf{S}$ in Section \ref{ssec:ft} with $w=1$ and $b=0$.
The final form of loss is a weighted combination of $\mathcal{L}_\text{AP}$ and $\mathcal{L}_\text{AM}$ as below.
\begin{equation}
    \mathcal{L}=(1-\alpha)\,\mathcal{L}_\text{AP}+\alpha\,\mathcal{L}_\text{AM}
\end{equation}
where $\alpha$ is between 0 and 1. For more than two speakers, although the ideal affinity matrix where embeddings from different speakers are opposite to each other cannot be achieved, the main aim for the AM loss is to increase the cosine distance for different speakers. Threhsolding in the next subsection resolves this issue.

\subsection{Absolute thresholding}
\label{mse_thres}

\begin{figure}[t]
\begin{minipage}[b]{1.0\linewidth}
  \centering
  \centerline{\includegraphics[width=6.8cm]{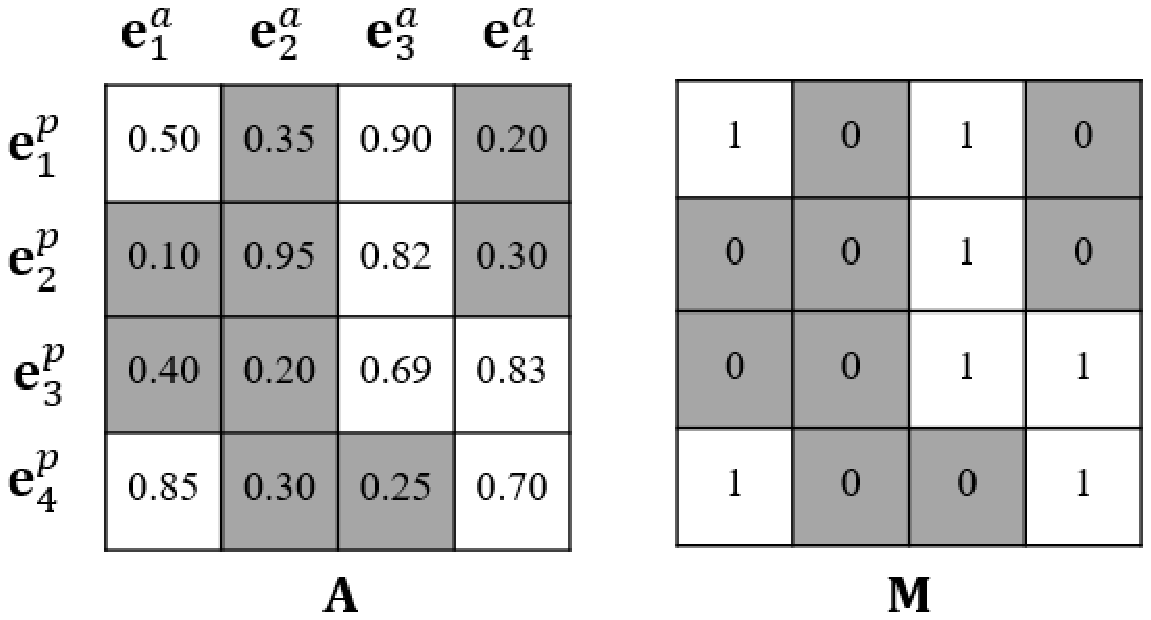}}
\end{minipage}
\caption{An example of $\mathbf{A}$ with its corresponding mask $\mathbf{M}$. The numbers in $\mathbf{A}$ are the cosine similarity scores between the speaker embeddings of the corresponding columns and rows. The AM loss is only calculated for cells that are not greyed out.}
\vspace{-0.3cm}
\label{fig:mask}
\end{figure}

For the row-wise thresholding step in spectral clustering, any value less than the threshold was set close to zero. Hence, it is important to encourage pairs of embeddings from different speakers to have a similarity score below the threshold during training. Therefore, a threshold, $t$, between 0 and 1 was used during training. The threshold can be applied for both the AM and AP losses.
For the AM loss, a mask matrix $\mathbf{M}=(\mathbf{M}_{i,j})\in\mathbb{R}^{N\times N}$ is created to only calculate the positive pairs whose similarity scores are lower than $t$ and the negative pairs whose similarity higher than $t$. That is,
    \begin{equation}
    \mathbf{M}_{i, j} = 
    \begin{cases}
        1 & \textrm{if } i = j, \mathbf{A}_{i,j} \leq t \\
        1 & \textrm{if } i \neq j, \mathbf{A}_{i,j} \geq t \\
        0 & \textrm{otherwise}
    \end{cases}
    \label{eq:absthr}
    \end{equation}
Thus the AM loss becomes
\begin{equation}
    \mathcal{L}_\text{AM} = \frac{1}{\sum_{i,j} \mathbf{M}_{i, j}}\sum_{i=1}^{N} \sum_{j=1}^{N}
    \mathbf{M}_{i, j}(\mathbf{I}_{i,j} - \mathbf{A}_{i,j})^2.
\end{equation}
The thresholding concept with $t=0.8$ is illustrated in Fig. \ref{fig:mask}. This enables the loss to focus only on the difficult positive and negative pairs as the ``easy negative pairs'' would have been thresholded to 0. 
Similarly, thresholding can be imposed on the AP loss, where a mask can be created such that the cross-entropy is calculated for all positive pairs and the difficult negative pairs, as the positive pairs are always needed. 

\subsection{Relative thresholding with Gaussian blurring}
\label{rel_thres}
\begin{figure}[t]
\begin{minipage}[b]{1.0\linewidth}
  \centering
  \centerline{\includegraphics[width=8.8cm]{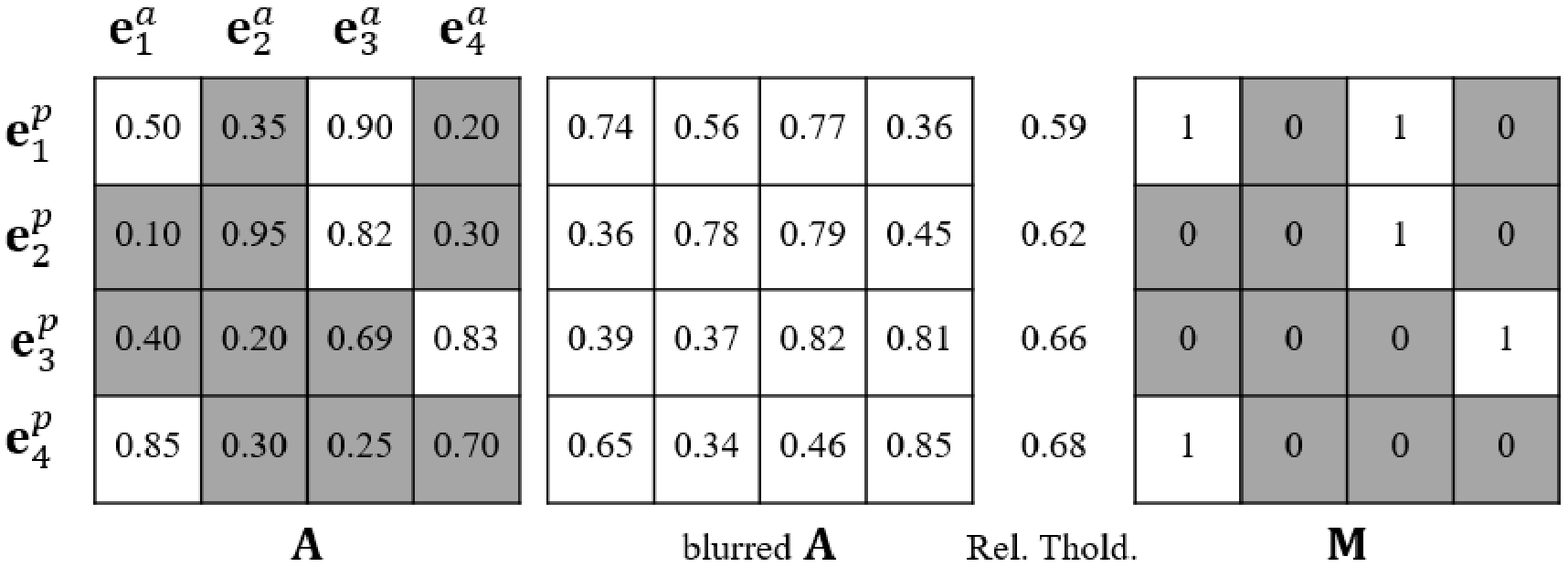}}
\end{minipage}
\vspace{-0.5cm}
\caption{$\mathbf{A}$ and $\mathbf{A}$ with Gaussian blur applied. The row-wise relative thresholds are obtained as described in step 4, which are then applied to $\mathbf{A}$, giving the mask $\mathbf{M}$. Rel. Thold. stands for Relative Threshold.
}
\vspace{-0.3cm}
\label{fig:rel_thold}
\end{figure}

As relative thresholding was used in spectral clustering, to closely simulate the row-wise thresholding during clustering, the Gaussian blur and relative thresholding are integrated into speaker embedding extraction model training.
This improves the estimation of the actual threshold by using Gaussian-blurred matrices.
To achieve this, the affinity matrix is first blurred to determine the relative threshold. The threshold which is in general different for different rows is then used on the unblurred affinity matrix to avoid gradient annihilation. 
The detailed steps are given below and shown in Figure \ref{fig:rel_thold}.
\begin{enumerate}
    \setlength\itemsep{0em}
    \item Compute the affinity matrix.
    \item Replace all diagonal elements with 1 since, during clustering, diagonal elements are the maximum of the row.
    \item Perform Gaussian blurring on the affinity matrix.
    \item Determine the relative threshold of each row by multiplying $t$ with the value of the diagonal element.
    \item Use the relative threshold to create a single mask matrix for both of the AP and AM losses. 
\end{enumerate}

\section{Experimental Setup}
\label{sec:exp}

\subsection{Data}
The dataset used to fine-tune and evaluate the final speaker diarisation performance was the AMI meeting corpus \cite{ami}, which consists of meeting recordings with 4-5 speakers per meeting. The training set contains 135 meetings with 155 speakers. The development (Dev) and evaluation (Eval) sets followed \cite{gs}.
Moreover, the joint VoxCeleb1 and VoxCeleb2 data were used as an intermediate fine-tuning stage followed by the final fine-tuning on the AMI training set in a two-stage fine-tuning setup. 
The input features of all models were beamformed raw waveform with BeamformIt \cite{beamformit}.

\subsection{System specifications}
During training, a 2-second window was sampled from each speech segment as input. During inference, speaker embeddings were extracted from the penultimate layer of the embedding extraction model. An average pooling layer was added on top of the wav2vec 2.0 model to produce an embedding for each window. This was followed by two additional fully connected layers, the first layer projected the embedding to the desired dimension (128D) and the second one was used to classify each input into a speaker label. 
The wav2vec 2.0 encoder was frozen for the first 10\% of fine-tuning steps and a triangular learning rate scheduler was adopted. For two-stage fine-tuning, SCALE was only applied to the stage on AMI as there was a mismatch between the nature of the VoxCeleb and AMI data.
For the baseline model, the angular Softmax (A-Softmax) loss \cite{Huang2018AngularSF, sphereface} was used for speaker classification, with $m=1$. 

During inference, each speaker-homogeneous segment was split into 2-second windows with a 1-second overlap. Speaker embeddings were obtained for each window and spectral clustering described in Sections~\ref{sec:sc} and \ref{sec:scale} were applied. All windows in a segment were assigned the same speaker label, following \cite{gs}.
The hyperparameters for spectral clustering (the $p$-th percentile and standard deviation for Gaussian blur) were tuned on the Dev set by grid search and applied to the Eval set directly. Three baselines were used, including a non-wav2vec \emph{TDNN baseline} following \cite{gs}, a \emph{classification baseline} using A-Softmax loss for fine-tuning wav2vec 2.0 and the strongest \emph{AP loss baseline} that used the AP loss for fine-tuning.

\subsection{Evaluation}
Both reference and automatic segments were used for evaluation.
Automatic segments were found using the same VAD and CPD in \cite{gs}. For the reference segmentation, the SER was scored with a 0.25-second collar on both sides of the segment without overlap. For the automatic segmentation, DER which is the sum of the SER, the missed speech (MS) and false alarm (FA), was reported. To avoid scoring against a large amount of long silence in the original reference, diarisation results with automatic segmentation were scored against the modified reference following \cite{gs}. In addition, a meeting-by-meeting sign test was performed to show the statistical significance of any improvements where appropriate.

\section{Results and Discussion}
\label{sec:res}

\vspace{-0.1cm}
\subsection{With one-stage fine-tuning}
\vspace{-0.1cm}
In this part of the experiment, wav2vec 2.0 was directly fine-tuned on AMI with SCALE and the results are shown in Table \ref{table:res_mod32}. 
For SCALE, $\alpha$ was set to 0.5 for the AP loss, and $t$ was set to $0.8$ and $0.95$ for the absolute and relative thresholding respectively. For relative thresholding with Gaussian blur, $\sigma$ was set to 1. 
\begin{table}[t]
\centering
\begin{tabular}{lcc}
\toprule
System & \multicolumn{2}{c}{SER (\%)} \\
& Dev & Eval \\
\midrule
TDNN \cite{gs} & 14.3 & 15.4 \\
Classification with A-Softmax & 9.6 & 19.3 \\
AP loss & 9.9 & 17.1 \\
\cmidrule{1-3}
AP loss + AM loss & 8.8 & 16.4 \\
AP loss + AM loss (Abs. Thold.) & 7.5 & 15.8 \\
AP loss (Abs. Thold.) + AM loss (Abs. Thold.) & \textbf{6.6} & 13.2 \\
AP loss (Rel.~ Thold.) + AM loss (Rel.~ Thold.)  & 6.9 & \textbf{12.0} \\
\bottomrule
\vspace{-0.5cm}
\end{tabular}
\caption{SER on AMI Dev and Eval sets using systems with SCALE fine-tuned on AMI only. TDNN is a non-wav2vec baseline while all the other systems were fine-tuned wav2vec 2.0. Abs. Thold. refers to the absolute thresholding (Sec. \ref{mse_thres}) and Rel. Thold. refers to the relative thresholding (Sec. \ref{rel_thres}) for all tables.}
\label{table:res_mod32}
\end{table}
%
The two wav2vec 2.0 baselines performed better on the Dev set while worse on the Eval set compared to the TDNN baseline\footnote{A clear comparison with the literature is hard as various setups are used, e.g. VBx\cite{vbx} gives good error rates but cannot be directly compared.}

With the AM loss term added,
the Dev and Eval sets SERs reduced, demonstrating the effectiveness of the AM loss term. Next, when absolute thresholding was applied to the AM loss, reductions in SERs on both Dev and Eval sets were found and the Eval set SER was similar to the TDNN baseline while the Dev set DER is clearly lower. Finally, by applying the SCALE method with absolute thresholding on both AM and AP losses, the system achieved 27\% relative SER reductions on both the Dev and Eval sets compared to using only the AP loss. 

Applying thresholding on both losses was more effective than only applying it on the AP loss, as the training process put more emphasis on the harder pairs. When using a relative threshold, the Dev SER was slightly higher than when using an absolute threshold, but the Eval SER was lower. 
A meeting-level sign test showed that the improvement achieved by SCALE compared to the classification baseline was statistically significant, and there was no statistical difference between absolute and relative thresholding results. 
\vspace{-0.1cm}
\subsection{With two-stage fine-tuning}
\vspace{-0.1cm}
\begin{table}[t]
\centering
\begin{tabular}{lcc}
\toprule
System & \multicolumn{2}{c}{SER (\%)} \\
& Dev & Eval \\
\midrule
Classification with A-Softmax & 9.0 & 14.7 \\
AP loss & 7.0 & 9.6  \\
\cmidrule{1-3}
AP loss + AM loss & 6.1 & 9.7 \\
AP loss + AM loss (Abs. Thold.) & 6.4 & 11.2 \\
AP loss (Abs. Thold.) + AM loss (Abs. Thold.) & \textbf{3.3} & \textbf{4.7} \\
AP loss (Rel.~ Thold.) + AM loss (Rel.~ Thold.) & 3.4 & 4.9 \\
\bottomrule
\vspace{-0.5cm}
\end{tabular}
\caption{SER on AMI Dev and Eval sets using systems first fine-tuned on VoxCeleb1$+$2, and then on AMI with SCALE. All systems used fine-tuned wav2vec 2.0.}
\label{table:res_mod32_vox_ami}
\end{table}
In the two-stage fine-tuning, SCALE was applied to the final fine-tuning on AMI, where $\alpha=0.5$ and $t=0.9$ were used for both absolute and relative thresholding. For relative thresholding, $\sigma=0.5$ was used with Gaussian blur. The results are shown in Table \ref{table:res_mod32_vox_ami}. As before, fine-tuning with the AP loss resulted in better diarisation performance than with the classification loss. 
When applying SCALE with thresholding on both the AM and AP losses, the model performed much better than the baselines. 
Meanwhile, the gap between the Dev and Eval SER became smaller, and it was also discovered that embeddings with SCALE had a much smaller SER change during hyperparameter tuning for spectral clustering. Therefore, SCALE achieved better robustness to spectral clustering hyperparameters. Overall, the best system was achieved using absolute thresholding on both losses, resulting in a relative SER reduction of 53\% and 51\% on Dev and Eval sets respectively when compared to the AP loss baseline. As before, the improvements were found to be statistically significant, and there was no statistical difference between the two thresholding methods.


\vspace{-0.1cm}
\subsection{With automatic segmentation}
\vspace{-0.1cm}
To investigate the effectiveness of using SCALE to fine-tune a wav2vec 2.0 in the full speaker diarisation pipeline, experiments were performed with automatic segmentation obtained after VAD and CPD, with results shown in Table \ref{table:res_cpd}. The MS and FA rates obtained after VAD and CPD were 1.2\% and 4.0\% respectively for the Dev set, and 1.3\% and 3.6\% respectively for the Eval set. 
The system trained with AP loss only, and SCALE systems with absolute and relative thresholding on both losses were investigated, 
which were the same as those in Table \ref{table:res_mod32_vox_ami}.
\begin{table}[t]
\vspace{0.1cm}
\centering
\begin{tabular}{l c c}
\toprule
System & \multicolumn{2}{c}{DER (\%)} \\
& Dev & Eval \\
\midrule
TDNN \cite{gs} & 12.6 & 15.6 \\
AP loss & 14.1 & 17.1  \\
\cmidrule{1-3}
AP loss (Abs. Thold.) + AM loss (Abs. Thold.) & \textbf{9.9} & \textbf{10.1} \\
AP loss (Rel.~ Thold.) + AM loss (Rel.~ Thold.) & \textbf{9.9} & 11.2 \\
\bottomrule
\vspace{-0.5cm}
\end{tabular}
\caption{DERs on AMI Dev and Eval sets using automatic segmentation. MS + FA rates was 5.2\% and 4.9\% on the Dev and Eval set.}
\label{table:res_cpd}
\end{table}
Compared to the AP loss, using SCALE with absolute thresholding achieved the best performance among systems, 
with a relative reduction of 30\% an the Dev set and 40\% on the Eval set.

\section{Conclusions}
\label{sec:conc}
This paper proposed SCALE, spectral clustering-aware learning of embedding framework to fine-tune the pre-trained speaker representations for speaker diarisation. SCALE used a combination of the angular prototypical (AP) and the affinity matrix (AM) losses to learn the structure of the embedding space, and to reduce the mismatch between the speaker embeddings and spectral clustering. SCALE also includes the $p$-percentile thresholding and Gaussian blur of spectral clustering into training to further reduce the mismatch. Experiments on the AMI data with SCALE achieved a relative SER reduction of 53\% and 51\% on the Dev and Eval sets respectively, compared to the AP loss baseline using oracle segmentation. SCALE also achieved 30\% and 40\% relative DER reduction on the Dev and Eval sets respectively with automatic segmentation.

\newpage

\printbibliography

\end{document}